# MORPHOLOGICAL NUMBER-COUNTS FROM ULTRADEEP HST IMAGES

S. P. Driver, R. A. Windhorst
*Arizona State University, Tempe, USA*

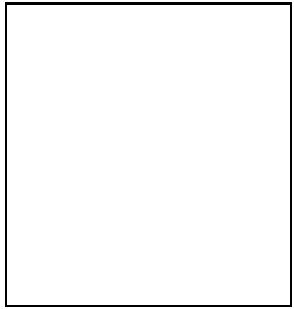

**Abstract**

We discuss the observations leading to the Faint Blue Galaxy problem and the uncertainties upon which faint galaxy models are based. Using deep Hubble Space Telescope (HST) imaging with the Wide Field Planetary Camera (WFPC2), we show how morphological information has been used to shed new light on the problem. Initial results indicate that the giant galaxies (ellipticals and early-type spirals), are well fit by *no-evolution* standard models down to $m_I \sim 24.5$ ($z \sim 0.8$). The data also show that the faint blue galaxies have late-type/irregular morphologies and cannot be adequately modelled until better constraints are placed on the local space density for this class of galaxies.

## 1 Faint Galaxy Number-counts

One of the most basic astronomical observations is to simply count the number of galaxies in a given direction as a function of apparent magnitude and use these observations to help understand the nature of our Universe. The advantages of such a simple observation, is that good statistical data can be obtained over a large magnitude range. Figure 1 shows some of the *number-count* data, obtained by many groups over a wide range of magnitude. The original contention [21], was that data such as these would allow a direct measurement of the cosmological parameters ($\Omega$, $q_o$, $\Lambda$) — the idea being that the departure of the number-counts from a purely Euclidean slope of 0.6 is attributable to the geometry of the Universe. The lines shown in Figure 1, reflect the predicted number-counts based on the best non-evolving standard models and it is clear that the models and the data strongly disagree at progressively fainter magnitudes. This discrepancy gives rise to the fairly long standing Faint Blue Galaxy (FBG) problem [23], [2], [24], discussed in the next section. In this article, we briefly outline the current state of the observations, the uncertainties that go into generating a faint galaxy model, and how the latest HST images, obtained with the WFPC2, now allow us to use morphological information to simplify the problem.

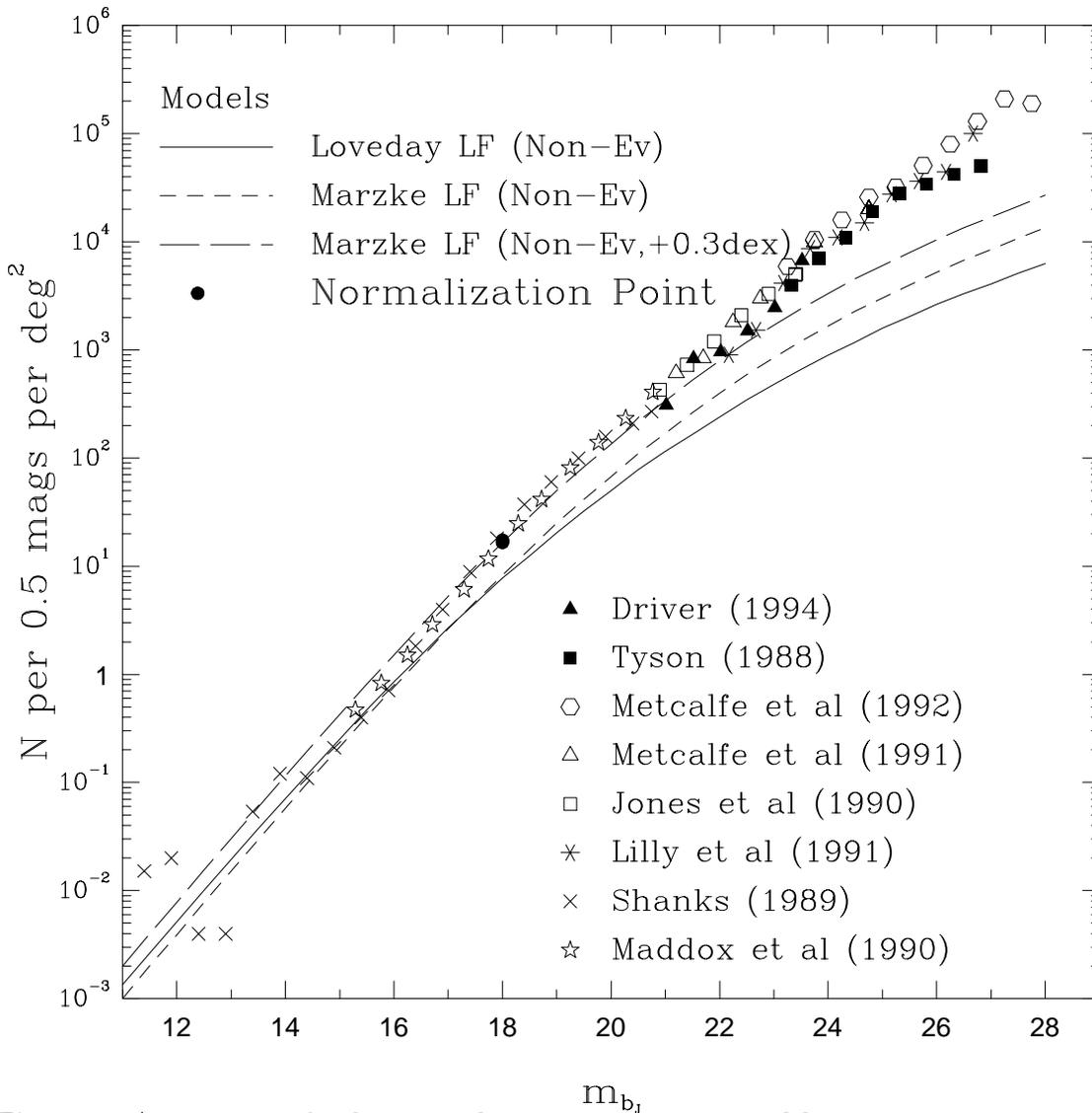

Figure 1: A montage of galaxy number-counts as measured by numerous groups over the past decade. The model lines show various predictions from the standard no-evolution models.

## 1.1 Observations

*1.1.1 Number-counts* Figure 1 shows the observed B-band number-counts, counts at longer optical and infra-red wavelengths follow shallower slopes at comparative magnitudes and come closer to matching the model predictions [24]. In fact the more recent deep K-band counts, [4], [16], lie close to the current standard model expectations. Essentially this simply states that the galaxies responsible for the faint excess in the B-band must be blue.

*1.1.2 Colours* The review article by Koo & Kron [24] contains an excellent diagram (see their Figure 2) of the colour-magnitude trend, from which two distinct pieces of information can be drawn. Firstly, the *mean* galaxy $(b_J - r_F)$ colour shifts towards the blue at fainter magnitudes (as we know it must from the number-counts in these bands), and secondly, the colour distribution broadens significantly at fainter magnitudes. This broadening is greater than the photometric errors at faint magnitudes, and would seem to imply that the observed galaxy population at faint magnitudes is *more diverse* than that observed locally in a magnitude limited sample. The mean observed colours of faint field galaxies in $(B - R)$, $(B - V)$ and $(V - R)$ have also been shown [8] to be consistent with the observed colours of late-type spirals and irregulars.

***1.1.3 Redshift surveys*** The ideal data set required to fully resolve the FBG problem is a large, complete and well defined redshift survey to faint magnitudes. Unfortunately measuring redshifts for faint field galaxies is a particularly time-consuming process and current resources restrict the measurement of redshifts to galaxies brighter than $m_b < 24$, where the problem is less distinct (c.f. Fig. 1). The redshift surveys that have been made, e.g. [2] [17], all point to a redshift distribution whose shape is consistent with that expected from the current standard no-evolution models, but of course inconsistent in terms of absolute numbers. In total, these surveys represent only a few hundred galaxies with incompleteness limits in the range 70-90 %. While it may seem unlikely that the missing galaxies are mostly at high or low-z, the possibility cannot be ruled out, as the selection biases are poorly defined, and are likely to be a strong function of spectral range sampling, surface brightness and epoch[1].

***1.1.4 Normalisation*** More worrisome perhaps, than the discrepancies between faint galaxy models and faint galaxy observations, is the apparent discrepancy between models and observations at relatively bright magnitudes ($m_b \sim 18$ mag). The solid line, in Figure 1, shows the current conventional model, based on a standard flat cosmology ($\Omega = 1, q_o = 0.5$) and the Mt. Stromlo-APM luminosity function (LF) [25]. And, at $m_b \sim 18$ mag ($z \sim 0.1$), there is already a discrepancy of a factor of two (0.3 dex) between the model and the galaxy counts. This discrepancy has been known for some time and is normally explained away in terms of a large local inhomogeneity [31], significant local evolution [26], incompleteness in the local surveys due to surface brightness and visibility constraints, [35] [5] [13], and/or systematic magnitude errors in the local surveys [28]. However, it is still unclear as to which effect(s) are responsible. Although, recent UKIRT observations, from which the local K-band LF has been derived [18], find a *higher* local normalisation suggesting that the previous local surveys have indeed suffered from high incompleteness.

## 1.2 Faint Galaxy Models

Despite the original aim of faint galaxy counts, the model predictions depend more critically on knowledge of the local LF [23] and any evolutionary processes [33], than on the cosmological parameters ($\Omega$ & $\Lambda$)[2]. To understand these dependancies, Figure 2 shows the inner workings of a typical faint galaxy model [7]. The thick line represents a prediction of the galaxy number-counts, based on knowledge of the local LF, k-corrections, a standard flat cosmology and no-evolution. Each of the dashed lines represents the contribution to the total counts from a narrow luminosity class. Immediately apparent are the effects of cosmology, most notably the k-corrections, which cause the departure of the individual lines from the Euclidean slope of 0.6. Hence at faint magnitudes the intrinsically more luminous (and therefore more distant) galaxies are effected more severely, and so the predicted counts at progressively fainter magnitudes depend more and more heavily on the contribution from the lower luminosity classes. Furthermore, at fainter magnitudes no one single luminosity class dominates the total counts, explaining perhaps the apparent diversity seen in the colour distribution and low correlation amplitude. Note that steepening the local LF has the effect of decreasing the distance between the dashed lines on Figure 2, so exasperating this dependancy. Likewise evolutionary effects may cause an independent steepening or flattening of each line, dependent on the level and type of evolution for that luminosity class. All in all, the number of potential unknowns is extremely large.

---

[1] A galaxy undergoing a recent starburst is likely to exhibit larger equivalent widths in the characteristic spectral lines and so be rendered more readily detectable.

[2] For $\Lambda = 0$ models, the predicted number-counts are relatively *insensitive* to the matter density [12].

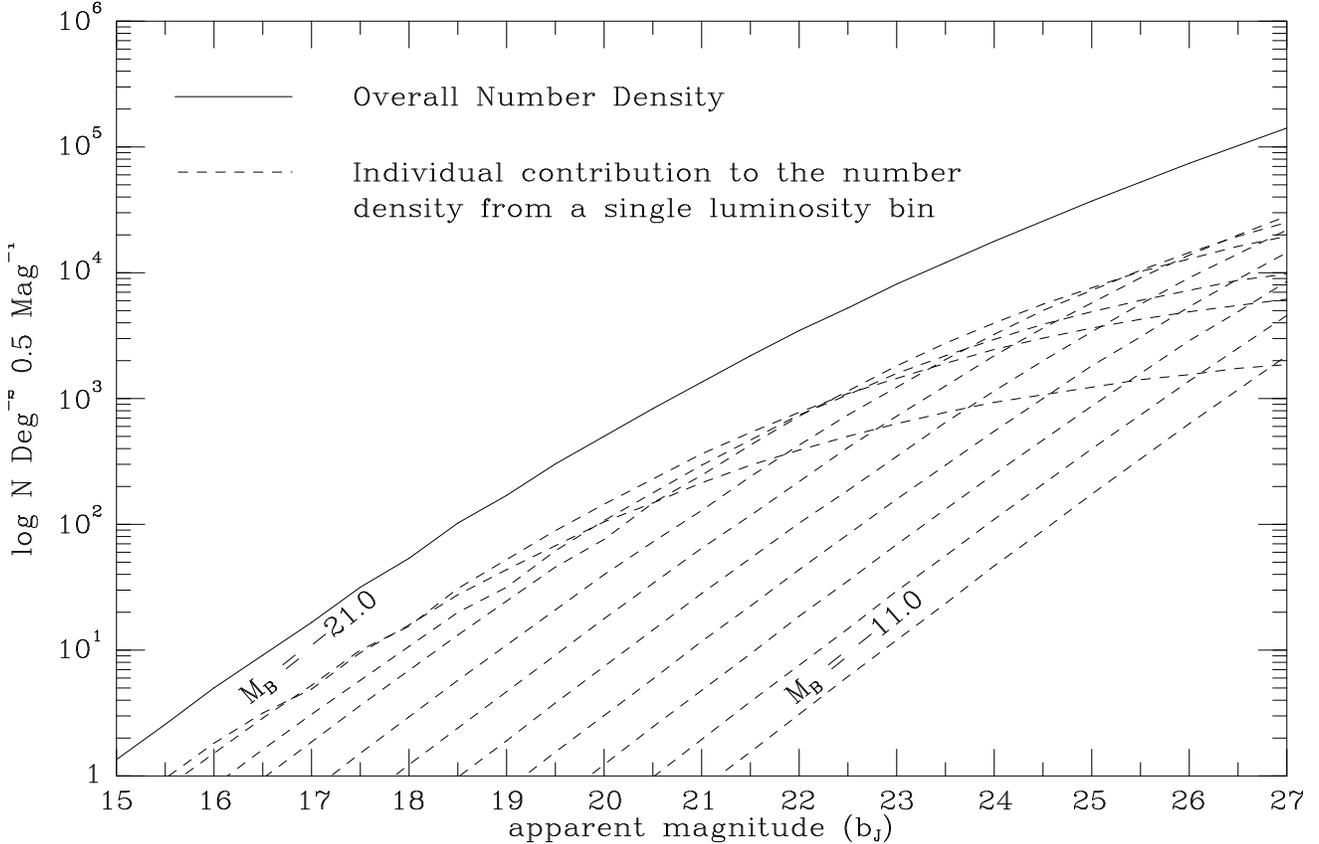

Figure 2: The "inner workings" of a faint galaxy model. The solid line shows the total prediction and the dashed lines show the contribution from each intrinsic luminosity interval.

*1.2.1 The Cosmological Parameters* Adopting an open $\Lambda = 0$ Universe (*i.e.* low $q_o$) reduces the decrease of the comoving volume element such that slightly higher counts are predicted at fainter magnitudes. However the gain is small, relative to the FBG problem (typically a factor of 2 at $m_b \sim 24$ compared to the deficit of 10 [12]). Alternatively adopting a large positive $\Lambda$ will increase the predicted numbers sufficiently, to begin reconciling the models with the data - however such a solution over-predicts the number-counts at longer observed wavebands [12] [15] [6].

*1.2.2 The Local Luminosity Function* A more basic dependency of faint galaxy models is upon the true local space density of galaxies [8]. Various groups, [13], [29], have postulated that surface brightness selection effects in the local surveys may have led to an underestimation of the slope of the true Schechter function (SFn) [30], which is typically used to describe the local space density of galaxies. A more subtle problem, however, is the inadequacy of using a single SFn to describe a wide range of luminosity and morphological types. One must be wary of extrapolating the bright end of a SFn to faint intrinsic luminosities, where the available local data is scarce. In particular, SFn's are incapable of following a sudden change or discontinuity in the local luminosity distribution.

To illustrate these points, Figure 3 shows the data from the Mt. Stromlo-APM survey [25] in which $\sim 1800$ galaxies were used to define a single global field SFn. The four insets represent four alternative luminosity functions. Models (a) and (b) are typical SFn representations, with differing slope parameters, while models (c) and (d) are Schechter-like functions, where an intial flat SFn is allowed to exhibit a discontinuous change in slope at some intermediate magnitude (intended to correspond to the giant/dwarf luminosity boundary at $M_{b_J} \sim -18$, [1]).

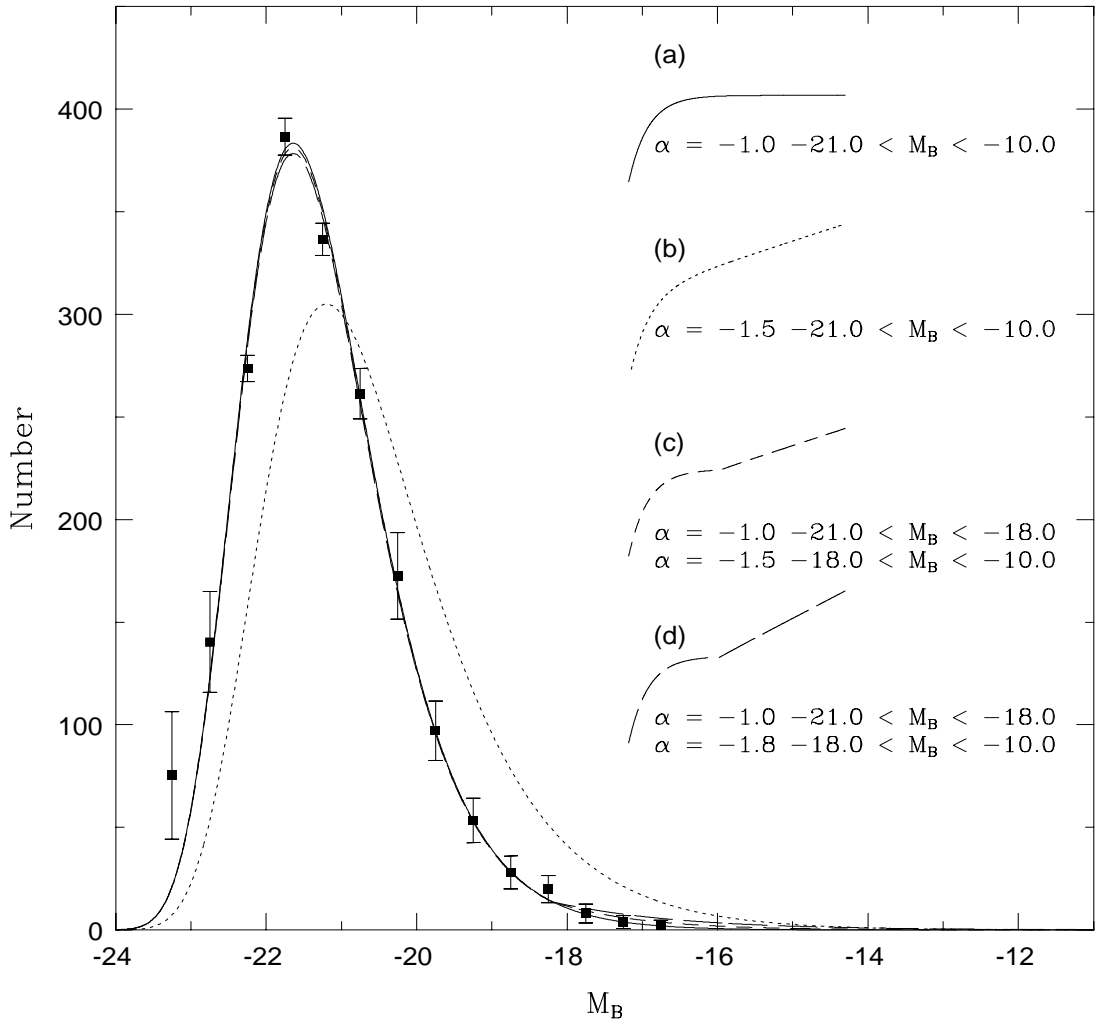

Figure 3: The observed Magnitude distribution of galaxies in a magnitude limited sample compared to four alternate LFs.

When these models are represented as they would be observed in a magnitude limited sample, *i.e.* in real numbers unadjusted for volume, models (a), (c) and (d) are indistinguishable. And, while the Mt. Stromlo-APM data can conclusively rule out a global SFn with a slope of $\alpha = -1.5$, it cannot rule out a turn up in the field luminosity function at luminosities fainter than $M_{b_J} \sim -18.0$. In reality this merely states that the local space density of low luminosity systems (dwarfs, irregulars etc) is extremely poorly constrained. One reason to perhaps suspect that the true luminosity distribution may indeed turn up, at fainter magnitudes, is from observations of clusters, where measurements of the luminosity distribution are more reliable. For example an upturn in the cluster LF has been noted in Coma, [20] [32], Virgo, [22], A963, [9], as well as various local groups [14], although some clusters show no such effect. In reality it is simply clear that there are no strong constraints, to date, as to the slope of the faint end of the field LF and, as already stated in §1.2, the predicted galaxy number-counts at faint magnitudes are critically dependant on knowing this faint end slope.

*1.2.3 Evolution ?* The more conventional explanation for the observed excess has always been to invoke some kind of luminosity evolution, [2] [33], and more recently merger evolution, fading dwarfs or other somewhat more imaginative scenarios. Clearly the Universe did not form with the current local galaxy population in place, and therefore some evolutionary process(es) must have occurred between the decoupling era and today. The question is when and by what mechanism. Typically luminosity evolution models are based on an initial star-burst, after which the galaxies luminosity decays typically exponentially with time. The resulting galaxy luminosities vary typically as $(1 + z)^\gamma$ and merger models normally follow the form luminosity $\propto (1 + z)^{-\gamma}$ and number density $\propto (1 + z)^\delta$. With reasonable values of $\gamma$ and $\delta$, the evolutionary effects far overshadow the discrepancy between low and high $q_o$ models. However with the current variety of locally observed galaxy types it is unlikely that any single evolutionary scenario is responsible, or if so, then galaxy evolution is most definitely not coeval or ubiquitous and any single evolutionary model is likely to fail.

## 2  Galaxy Morphology using the Hubble Space Telescope

As outlined in the previous section, the number of unknowns upon which the typical faint galaxy model is based are very large. Given these uncertainties, it seems hardly surprising that the models and observations are not well matched. Additional observational constraints are required to help simplify the problem at hand. Since the refurbishment of the HST, detailed images of faint field galaxies at $0.1''$ resolution have now be obtained. Typically previous ground-based images have been limited by atmospheric seeing to a resolution of $0.8''$ and normally worse in the case of multiple stacked images taken over an appreciable time-span. The HST offers a factor of almost 100 increase in the number of independently sampled pixels, and so allows morphological details to be seen down to the faintest detectable magnitudes. Indeed, the effects of cosmology assist in the sense that the angular diameter of a fixed object varies little beyond redshifts $z > 0.5$, such that it is an objects apparent surface brightness which limits the ability to discern morphology, rather than telescope resolution.

With morphological information, it becomes possible to simplify the modelling process. For example, galaxies of a similar type are more likely to have evolved via a similar evolutionary mechanism, and so it would seem more realistic to adopt a separate evolutionary model for each type. The local LF is also better established for galaxies of certain type, e.g. ellipticals and early-type spirals. Ultimately, morphological segregation allows each morphological type to be modelled independently, so reducing the complexity of each individual model.

### 2.1  The HST Morphological Number Counts

Recently four morphological surveys of faint field galaxies have been made from the HST Medium Deep Survey (MDS): CRGINOW [3], DWG [10], DWOKGR [11] and GES [19]. The CRGINOW survey is based on the entire WF/PC data-set (11,500 galaxies) and the morphological typing is automated to distinguish between disk and bulge systems. The WF/PC data covers the range $18 < m_I < 21$. The other three WFPC2 surveys (each containing between 150 and 300 galaxies) extend this range to $m_I < 24.5$, based on eyeball classifications, typically made by a number of independent observers and using both the image data and the objects major-axis surface brightness profiles. Various consistency checks suggest that the morphological accuracy is $\sim 1$ Hubble-type, hence the three WFPC2 studies classify galaxies into three broad classes, E/S0's (bulges), Sabc's (bulge-disks) and Sd/Irr's (disks and asymmetrical systems).

Figure 5 shows the morphological number-counts for (a) all types, (b) E/S0s, (c) Sabc's and (d) Sd/Irrs. The counts for all types, agree well with groundbased counts and show the usual trend of a steep slope. The dotted and dashed lines in Fig. 5a illustrate the conventional no-evolution models which show the usual underprediction of the galaxy counts at fainter magnitudes. The other three panels in Figure 5 represent the new morphological number counts, as revealed by MDS studies of WF/PC and WFPC2 data.

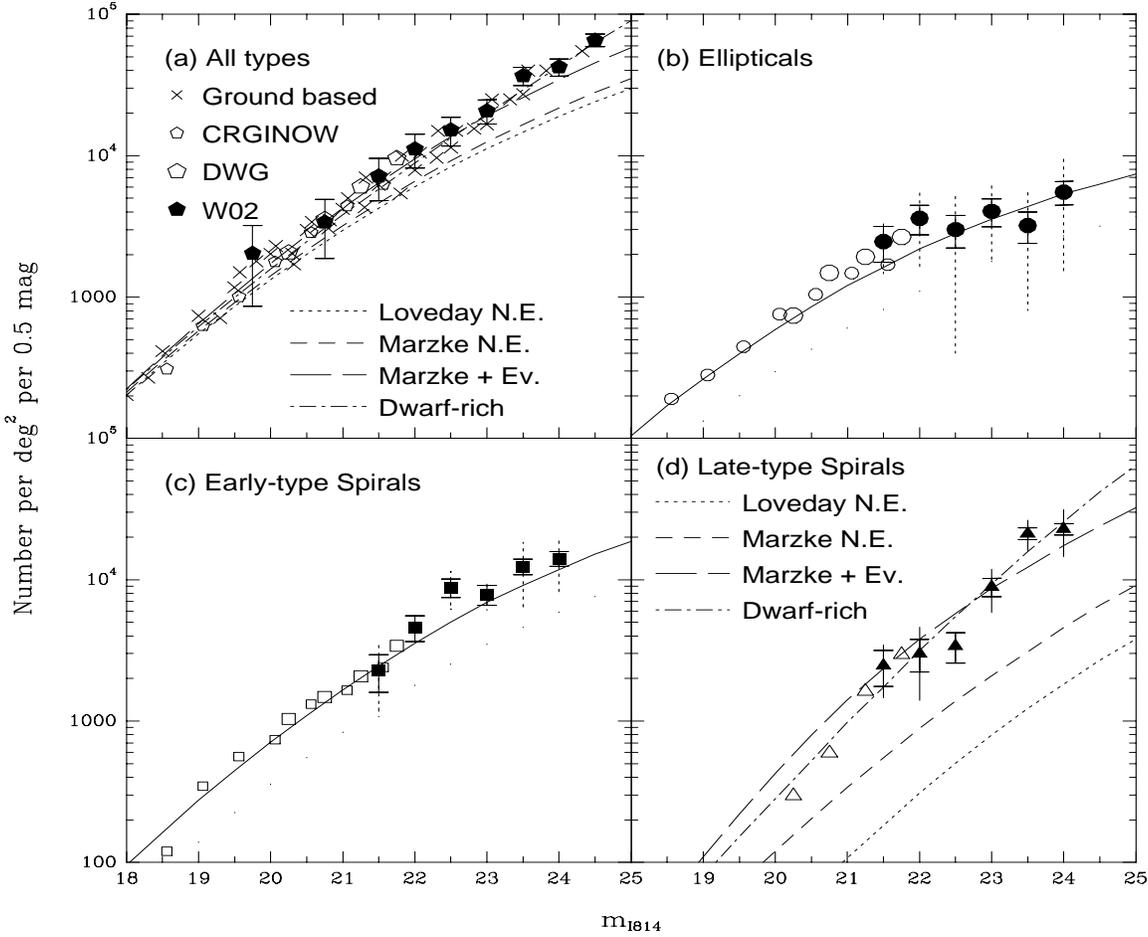
Figure 4: Morphological galaxy counts from HST WFPC2 images

## 2.2 Modelling the Morphological Number-counts

For each of the morphological groupings shown in Figure 5b, c & d we show the predictions from the standard best fit no-evolution models. These models assume a standard flat cosmology and the local LFs as derived by MGHC [27], all re-normalised by the same factor of 2 for the reasons outlined in §1.1.4. For simplicity, no-evolution is incorporated into the models and the models do not include any isophotal selection effects. From comparisons between the predictions of these basic models and the morphological counts, we can see that:

(i) Elliptical and S0s appear to follow closely the prediction based on a standard flat cosmology, no-evolution and a flat re-normalised luminosity function.

(ii) Early-type spirals similarly appear to follow the predicted number-counts for this type.

(iii) The Late-type Spirals and Irregulars are severly under-predicted by the models based on either the Loveday *et al* LF [25], or the Marzke *et al.* LF [27].

The primary conclusion is clear: *it is the late-type/Irregular population which is responsible for the faint blue galaxy excess*, and this is fully consistent with the earlier discussion that it is the properties of these types of galaxies of which we are most uncertain. The second conclusion is more unexpected in that, *the elliptical and early-spirals are remarkably well fit by the standard no-evolution models*. This implies that little or no evolution has occurred in these populations out to $m_I \sim 24.25$ (or $z \leq 0.8$), and that the standard cosmological model cannot be grossly incorrect. In fact, the possibility now exists to return to the initial aim of faint galaxy number-counts and start using morphological galaxy counts, of E/S0's say, to constrain cosmological models, and in particular $\Lambda$. This exciting prospect is not discussed further here but see [12] for more details.

Two more subtle points are also borne out with regards to LF normalisation. The first is that the individual LFs have each been re-normalised by the same amount, the fact that despite this both the E/SO and Sabc models match their respective counts, argues against an evolutionary or incompleteness explanation to the normalisation problem — as such solutions would be expected to effect these types to different degrees. The second point is that the shape of the E/SO and Sabc curves closely follow the shape of the counts, regardless of normalisation. Hence, if significant local luminosity or number evolution had occurred, then it would have to switch off, over the regime shown in Figure 5, for the shape of the counts to remain consistent with the no-evolution predictions. Both of these points are somewhat tentative, but argue for a large local inhomogeneity or a systematic zero point error explaining to the normalisation problem.

Finally, concentrating on the late-type Irregular population alone, we can attempt to determine how best the models can be reconciled with the observations. If these systems are genuine late-type spirals and Irregulars, their intrinsic luminosities are likely to be low, and hence their corresponding redshifts will also be low, ruling out any cosmological effects. Briefly we can consider the two extremes: (a) a non-evolving solution, and (b) an evolving solution based on the conventional Mt. Stromlo-APM LF.

*2.2.1 A Dwarf Dominated Model* To reconcile the predicted number-counts to the observations of Sd/Irr's alone without evolution requires increasing the normalisation of the late-type LF by a factor of 5 and increasing the Schechter slope to $\alpha = -1.8$ (dashed line on Fig. 5d, [8]). Such an increase has been shown to be inconsistent [7] with the faint redshift surveys, e.g. [17], which find little or no evidence for a low-z bump in the redshift distribution at faint magnitudes.

*2.2.2 An Evolving Standard-Dwarf Model* Conversely we can consider the level of evolution required to reconcile the observations to the predictions by adopting some simple parmaterisation of the luminosity evolution of dwarf galaxies, assuming the conventional flat faint end slope [25]. For example, if we assume a ubiquitous starburst at z = 0.5 in the entire population, afterwhich the luminosity decays exponentially with time [29], we find that 2 magnitudes of evolution is required in the entire dwarf population to match the observations ([11], large dashed line on Figure 5d). Although alternative evolutionary scenarios are possible, such as merging, the level of evolution required is extraordinarily high.

# 3 Conclusions

The faint blue galaxy problem, which has been with us for almost two decades, has been "morphologically resolved" due to the superb quality of the WFPC2 data from the refurbished Hubble Space Telescope. By segregating the faint galaxy number-counts into three broad morphological types (E/S0, Sabc, Sd/Irr), a significant step forward in the observational constraints has been made. The independent elliptical and early-type spiral number-counts follow closely the predictions of the non-evolving standard model, based on a standard flat cosmology and knowledge of the local space densities of these types. No significant luminosity or merger evolution is required to reconcile these observations with the models and, assuming no cosmic conspiracy, this places the end of the epoch of giant galaxy formation to $z \geq 0.8$.

As a side product, it may now become possible to constrain cosmological models using these morphological number-counts, although clearly a far larger statistical sample is required, as is spectroscopic confirmation of the morphological typing. The principle remaining problem then, is the required renormalisation of the models at relatively bright magnitudes.

The morphological types responsible for the faint blue excess have been shown to be systems consistent with late-type/Irregular morphologies. Their observed number-counts disagree with model predictions for this type by a factor of 10 ! However, it is precisely this population for which the local properties (e.g. local space density and star-forming mechanisms) are poorly understood, so that the FBG problem is simply a reflection of this fact. Neither steepening the local faint end slope of the LF, nor evolution of a flat faint-end can alone provide viable solutions. However a combination of these two effects can provide a plausible solution [29].

In summation the faint blue galaxy excess has been isolated, the morphologies determined, but the balance between star-bursts, mergers and/or a steep faint end slope is unclear, and liable to remain so until improvements have been made in our knowledge of the local late-population.

**Acknowledgements.** We are grateful to the other Medium Deep Survey team members and in particular Richard Griffiths, Stefano Casertano, Eric Ostrander and Karl Glazebrook and also to Steve Phillipps and William Keel for informative discussions.